\documentclass[english,twoside,a4paper,10pt]{article}

\usepackage[latin1]{inputenc}
\usepackage[T1]{fontenc}
\usepackage[english]{babel}
\usepackage{amsmath}
\usepackage{amsfonts}
\usepackage{graphicx}
\usepackage{subfigure}
\usepackage{a4wide}
\usepackage{amssymb}
\usepackage{fancyhdr}
\usepackage{mathrsfs}
\usepackage{color}
\usepackage{verbatim}
\usepackage[toc,page]{appendix}

\def\ba{\begin{eqnarray}}
\def\ea{\end{eqnarray}}

\def\be{\begin{equation}}
\def\ee{\end{equation}}

\def\p{\partial}

\begin{document}

\title{\bf{ Bounce Models within Teleparallel modified gravity}}

\author{ 
Alessandro Casalino$^{1, 2}$\footnote{E-mail address: alessandro.casalino@unitn.it},\,\,\,
Bruno Sanna$^3$\footnote{E-mail address: b.sanna1@studenti.unipi.it},\,\,\,
Lorenzo Sebastiani$^{4}$\footnote{E-mail address: lorenzo.sebastiani@pi.infn.it},\,\,\,
  Sergio Zerbini$^1$\footnote{E-mail address: sergio.zerbini@unitn.it}\\
\\
\begin{small}
$^1$Dipartimento di Fisica, Universit\`a di Trento, Via Sommarive 14, 38123 Povo (TN), Italy.
\end{small}\\
\begin{small}
$^2$TIFPA - INFN,  Via Sommarive 14, 38123 Povo (TN), Italy.
\end{small}\\
\begin{small}
$^3$Dipartimento di Fisica, Universit\'a di Pisa, Largo B. Pontecorvo 3, 56127 Pisa, Italy.
\end{small}\\
\begin{small}
$^4$Istituto Nazionale di Fisica Nucleare, Sezione di Pisa, Italy.
\end{small}
}
\date{}

\maketitle

\abstract{
In this paper, working in a Friedman-Lemaitre-Robertson-Walker (FLRW) space-time, we recover, without invoking unconventional exotic matter, the generalized Friedman equation of Quantum Loop cosmology, and therefore the cosmological bounce. We obtain this result proposing a new model in the framework of modified teleparallel gravity, where the Ricci scalar is replaced by the torsion scalar $T$. Furthermore we show that the associated perturbations again in a flat FLRW space-time are not affected by superluminarities or gradient instabilities. Then, we generalize the results to the curved FLRW space-time by using an appropriate formulation. In this context, the results of Born-Infeld model are also investigated.
}

\section{Introduction}

General Relativity (GR) with the presence of a suitable cosmological constant and the addition of cold dark matter, the so called $\Lambda$CDM model, is able to describe a large part of the cosmic history of our universe. The cosmic acceleration or dark energy dominated era which our universe undergoes today is well supported by the cosmological constant which drives the de Sitter expansion, and recently the $\Lambda$CDM model has been tested with high accuracy \cite{Plank1,Plank2}\footnote{Although we should mention that a tension associated with different measure of Hubble constant is present, and its currently under study, see for instance \cite{Yang:2018euj, Vagnozzi:2019ezj}. }.
%%%%%%%%%%%%%%%%%%%%%%%%%%%%%

However, it is well known that GR admits singular space-time solutions, where 
scalar curvature invariants become singular and geodesic incomplete metrics exist. In radiation/matter dominated universe the Big Bang singularity occurs and only under the assumption of
unconventional exotic matter it can be avoided and eventually replaced by a \textit{bounce}, where a contraction phase is followed by an expansion and the scale factor reaches a minimum but finite value (see Ref. \cite{bouncereview} for a review).
%%%%%%%%%%%%%%%%%%%%%%%%%%%%%%%%%%%%%%%%%%%%%%%%%%%%%%%%

In Quantum Loop Cosmology (QLC) an effective modified Friedman equation has been obtained and the solution in a flat Friedmann--Lema\^{i}tre--Robertson--Walker (FLRW) space-time admits the cosmological bounce \cite{L1,L2,L3,L4, aimeric}. Several modified gravity models, including mimetic gravity \cite{Muk},
non-polynomial gravity \cite{aimeric2},
Einstein Gauss-Bonnet \cite{nostri1}, Einstein-Aether (EA) gravity \cite{nostri2}, lead to loop modified Friedmann equation for flat FLRW space-time. 

Here, following Refs. \cite{nostri1, nostri2}, we consider a different framework with respect to GR. In particular we will analyze modified teleparallel gravity models. Teleparallel gravity is an approach to gravity in which, instead of making use of the Levi-Civita connection, the so called Weitzenbock connection \cite{torsion1, torsion2, torsion3} is considered. This choice of the connection leads to a vanishing curvature tensor, but to a non vanishing torsion tensor. 
In the modified teleparallel framework the Lagrangian of the theory is a function $f(T)$ of the torsion scalar $T$, which is a suitable combination of quadratic scalars depending on the torsion tensor. These theories involve second order differential equations of motion, making them less problematic than others where the modifications are proposed as general functions of curvature invariants, as in $F(R)$-models.

However we should not forget that teleparallel gravity is a variant of Riemann-Cartan geometry where a spin connection is present. In this paper we introduce the notion of \textit{proper frame} in which the spin connection is vanishing, and some issues related to the breaking of the local Lorentz invariance are discussed. 
We limit our discussion to highly symmetric space-times, namely FLRW space-times, where simple proper frames can be found. For additional discussions regarding the so called \textit{covariant teleparallel gravity} formalism, see for instance Ref. \cite{Sari16}.

We focus on solutions without singularities. In particular we recover the equations of QLC without considering exotic matter. We study two bounce models: one is the Modified Born-Infleld model \cite{FerraroBI}, while the second is a novel $f(T)$ proposal. We consider the aforementioned models in a spatially flat and curved FLRW space-time. In fact, the possible relevance of curved FLRW space-times has been recently pointed out in \cite{Dival,Vag}.

The paper is organized as follows. In Section {\bf 2} we briefly review the formalism of teleparallel gravity and the teleparallel formalism of GR. In Section {\bf 3}, we discuss a $f(T) $ model which reproduces the QLC results in a flat FLRW space-time. A Born-Infeld model is also revisited. The cosmological perturbations of these models are studied in Section {\bf 3.2}, where we review some results (mainly from \cite{FT43}) and show that the theory is not affected by superluminarity effects or gradient instabilities.  In Section {\bf 3.3} the curved case in investigated. Using the mini-superspace approach, the field equations of the theory are derived. Finally, conclusions are given in Section {\bf 4}.  

In our convention, the speed of light $c=1$ and the Planck mass $1 / 8 \pi G = 1$. We also adopt the mostly plus metric convention.

\section{Teleparallel gravity review}

In this section we review the Riemann-Cartan and Weitzenbock teleparallel geometry and the main difference with respect to textbook's GR. In fact, in GR one usually considers the Levi-Civita connection, here denoted by $\hat{\Gamma}^\mu_{\alpha \beta}$. This connection satisfies the metric compatibility condition, i.e. the associated covariant derivative of the tensor metric vanishes. Furthermore it is symmetric in the lower indices. As a result, the theory is torsion free. 

In general, a generic metric compatible connection, or Cartan connection, has a non trivial anti-symmetric part, and the  torsion tensor is not vanishing. In the following, we shall present a short review of the formalism we will use throughout the paper. 

The geometry which deal with a Cartan or spin connection is called Riemann-Cartan geometry. In this context, the dynamical variables are one-forms $e^a$, which constitute the tetrad (or vierbein in $d=4$), and the spin connection one-forms $\omega^a_b$. Introducing the natural basis one-form  $dx^\mu $ by means of $e^a=e^a_\mu dx^\mu$, and  $\omega^a_b=\omega^a_{b\mu} dx^\mu$, the metric tensor is related to the tetrads via
\be
g_{\mu \nu}=\eta_{ab}e^a_\mu  e^b_\nu\,,
\ee
where $\eta_{ab}$ is the Minkowski metric tensor of the tangent space. From this we note that the metric determinant which effectively appears in the Lagrangian is $\sqrt{-g}=det (e^a_\mu)$.

The related Riemann-Cartan and torsion two-forms are given by Cartan equations
\be
R^a_b=d\omega^a_b+\omega^a_c \wedge \omega^c_b\,,
\ee
\be
T^a=de^a+ \omega^a_b \wedge e^b\,.
\ee
If we denote by $L=L(x)$ a local Lorentz transformation, namely $\eta=L^T \eta L$, one has 
\be
e'=Le\,, \quad g'=g\,, \quad \omega'=L\omega B+LdB\,\quad  \text{where}\quad B=L^{-1}\,,
\ee
where $g$ is the metric tensor. If we impose a vanishing Riemann-Cartan two-form, we obtain the so called Weitzenbock geometry, and the related connection is called the Weitzenbock connection, here denoted by $\Gamma^\mu_{\alpha \beta}$. It is possible to show that the most general solution of $R^a_b=0$ is $\omega=LdB=-dL B$, namely
\be
\omega^a_b=L^a_c dB^c_b\,.
\label{inertial}
\ee
Thus, the non vanishing related torsion two-form reads
\be
T^a=de^a+ L^a_c dB^c_b \wedge e^b\,.
\ee
Recalling that $e^a=e^a_\mu dx^\mu$ and $\omega^a_b=\omega^a_{b\mu} dx^\mu$, we get the torsion tensor as
\be
T^\rho_{\mu \nu}=\frac{e^\rho_a}{2}\left(\partial_\mu e^a_\nu +\omega^a_{b\mu}e^b_\nu\right)- (\mu \leftrightarrow \nu)\,.
\ee
Given a Cartan connection, we can introduce the contorsion tensor $K^\mu_{\phantom{\rho}\alpha \beta}$ by means
\be
\Gamma^\rho_{\mu \nu}=\hat{\Gamma}^\mu_{\rho \beta}+K^\mu_{\phantom{\rho}\rho \beta}\,.
\ee
Therefore the definition of $K^\mu_{\phantom{\rho}\alpha \beta}$ from the torsion tensor is
\be
2K^\rho_{\phantom{\rho}\mu \nu}=T_{\mu\phantom{\rho}\nu}^{\phantom{\mu}\rho}+T_{\nu\phantom{\rho}\mu}^{\phantom{\mu}\rho}-T^\rho_{\phantom{\rho}\mu\nu}\,.
\ee
We also have
\be
\Gamma^\rho_{\mu \nu}=e^\rho_a\left(\partial_\nu e^a_\mu +\omega^a_{b\nu}e^b_\mu\right)\,.
\ee
At this point, we can choose $\omega^a_b$ in equation (\ref{inertial}) to be null, since this choice simplifies the solutions of the equations of motion. This choice clearly still leads to vanishing Cartan-Riemann curvature, and thus the Weitzenbock connection depends now only on the tetrad, and the related inverse $e^\rho_a$, namely
\be
\Gamma^\rho_{\mu \nu}=e^\rho_a\partial_\nu e^a_\mu \,.
\ee
Finally, we introduce the torsion scalar, the only  scalar, which being a particular combination of the three independent quadratic torsion scalars, leads to second order differential equations of motion, 
\be
T=T_{\rho\mu\nu}S^{\rho\mu\nu}\,,
\ee
where
\be
S^{\rho\mu\nu}=K^{\rho\mu\nu}+g^{\rho \mu}T^{\beta \nu}\,_\beta-g^{\rho \nu}T^{\beta \mu}\,_\beta\,.
\ee
\\
Lastly, a small remark regarding the choice $\omega^a_b=0$ is necessary. In fact this choice leads to a manifest breaking of the Local Lorentz Invariance (LLI), since $dT^a$ is no longer associated with a tensor 2-form.

In the context of the \textit{teleparallel equivalent of general relativity} such an arbitrary choice for the spin connection does not pose a severe problem, since the term in the torsion scalar depending on $\omega$ can be rewritten as a total derivative (see Ref. \cite{Krrsak17}), and thus does not contribute to the equations of motion derived from a Lagrangian of the form \ref{tegr}, which are still invariant under LLI. However, when working with generalizations like $f(T)$ gravity, if $f(T) \neq T$, this is no longer true, and the variation of such term will be in general not vanishing. Unless we want to use a frame independent formalism (see for example Ref. \cite{Sari16}), the only solution is to choose a frame, whose tetrad called \textit{proper tetrads}, where the term in the Torsion scalar proportional to the spin connection vanishes. In fact choosing a FLRW space-time leads exactly to this, thus there will be no contradictions in the following analysis. For more details about the problem of covariance of teleparallel gravity see Ref. \cite{Koiv, Hoh, Fio, Fer00, Fer01, Ferraro11}.

\subsection{Teleparallel formulation of General Relativity}

In this section we review the teleparallel formulation of GR. If we consider the Weitzenbock connection,  one has the important identity
\be
\sqrt{-g}\hat{R}=-\sqrt{-g}T-\partial_\mu(2 T^{\beta \mu}\,_\beta\  )\,,\label{eq:R}
\ee
where $\hat{R}$ is the standard GR Ricci tensor obtained with the Levi-Civita connection. Since the second term of equation \eqref{eq:R} is a surface term, the action 
\be
I=-\frac{1}{2}\int dx^4 \sqrt{-g}T +I_m
\label{tegr}
\ee
leads to equation of motion which are equivalent to the GR ones.

As an important example, consider the flat FLRW space-time. Here one is dealing with high symmetric space-time, namely 
\be
ds^2=-dt^2+a(t)^2 ds_0^2\,,
\ee
where $a(t)$ is the scale factor and depends on the cosmological time only. If we use the Cartesian spatial coordinates $x^i$, one has $d s^2_0=\delta_{ij} dx^idx^j$, and a suitable choice for the tetrad is the naive diagonal one, namely
\be
e^a_\mu=diag(1,a(t),a(t),a(t))\,.
\label{vierbFRW}
\ee
In this case we can safely use the Weitzenbock connection $\omega^a_b=0$, being (\ref{vierbFRW}) what we defined a \textit{proper tetrad}. The torsion scalar $T$ reads
\be
T=6H^2\,,
\ee
and the above teleparallel action leads to the GR Friedmann equation
\be
3H^2=\rho\,.
\ee

\section{Modified Teleparallel Gravity}

It is well known that a possible description of Dark Energy (DE) can be achieved by the modified gravity models based on $f(R)$-gravity, where $R$ is the Ricci scalar. However, this approach involves fourth order differential equations of motion, possibly leading to Ostrogradskij instabilities. In this section we want to discuss a similar theory, where the function is a function of the scalar torsion $T$ instead of the Ricci scalar. In fact, one may introduce the modified teleparallel gravity considering a Weitzenbock geometry with a Lagrangian depending only on the scalar torsion $T$, namely $f(T)$-gravity \cite{Tama, fT1, fT12, fT13, fT14, fT2, fT21, fT22, fT3, fT31, fT40, fT41, fT42, FT43}. A review on this approach can be found on Ref. \cite{SariRep}.

The action of these models is defined as
\be
I=-\frac{1}{2} \int dx^4 \sqrt{-g}f(T) +I_m\,.
\ee
It is possible to show that, with this choice of the tensor scalar $T$, the related equations of motion are second-order partial differential equations.

As an example, we verify this important feature of the $f(T)$ models considering again the flat FLRW space-time. In fact, in this relevant case, the Cartesian and spherical coordinates lead to the same expression for the scalar tensor, namely $T=6H^2$. Moreover, the generalized Friedmann equation is \cite{SariRep}
\be
\rho=\frac{f(T)}{2}-T f_T(T)\,,
\ee
or, using the value of the torsion tensor in this space-time $T=6H^2$,
\be
\rho=\frac{f(H)}{2}-H\frac{f'(H)}{2}\,,
\label{k0}
\ee
where $\rho$ is the standard matter energy density and $f'(H)=\frac{d f}{d H}$. The diffeormorphism invariance leads to
\be
\dot{\rho}=-3H(\rho+p)\,,\label{conslaw}
\ee
and the second Friedmann equation follows as usual by the above two equations (\ref{k0})--(\ref{conslaw}), namely
\be
-p=\frac{f(H)}{2}-H\frac{f'(H)}{2}-\frac{1}{6}\dot{H} f''(H)\,,
\ee
where the dot denotes the time derivative and $p$ the standard matter pressure.

\subsection{Cosmological bounce models}
\label{sec:cosmo_bounce_models}

In this section we propose two bounce models for the early Universe in the context of teleparallel gravity. For additional works about bounce models in teleparallel gravity, see \cite{delaCruz-Dombriz:2018nvt,Caruana:2020szx}.

\subsubsection{A model proposal}

We introduce the following expression for the function of the torsion scalar
\begin{equation}
f(T)=  \frac{12}{\alpha^2}\left[
 1-\sqrt{1-\alpha^2 \frac{T}{6}}-\alpha \sqrt{T/6}\arcsin\left(\alpha \sqrt{\frac{T}{6}}\right)
 \right]\,,\label{mod1}  
\end{equation}
which can be rewritten, using $T=6H^2$,
\begin{equation}
 f(H)=\frac{12}{\alpha^2}\left[
 1-\sqrt{1-\alpha^2 H^2}-\alpha H\arcsin\left(\alpha H\right)
 \right]\,,\label{eq:emp}
\end{equation}
where $\alpha$ is a dimensional positive parameter. Note that $f(H)=-6 H^2+ O(\alpha^2)$ when $|\alpha|\rightarrow 0$, confirming that we are studying a correction to Einstein gravity which can be controlled with the the value of the parameter $\alpha$.

The first Friedmann equation is
\begin{equation}
\frac{6}{\alpha^2}\left[
1-\sqrt{1-H^2\alpha^2}\right]=\rho \,,\label{EOM1bis}
\end{equation}
which is equivalent to the QLC Friemann equation
\begin{equation}
3H^2=\rho\left[1-\frac{\rho}{\rho_c}\right]\,,\qquad \text{where}\qquad  \rho_c=\frac{12}{ \alpha^2}\,,
\label{zf}
\end{equation}
where $\rho_c$ is called critical density. Thus, by assuming $p=\omega \rho$, with $\omega\neq -1$, when 
$ \rho_c \rightarrow \infty$, we recover Einstein's gravity. 
It is well know that the above equation leads to a cosmological bounce solution . For example the density reads \cite{nostri2}
\be
\rho=\frac{1}{\alpha^2+\frac{3(1+\omega)^2}{4}t^2}
\ee
As a result, the correction to GR avoids the Big bang singularity. In this case, the model admits a bounce solution with $H=0$. On the other hand, if $\omega=-1$, namely $\rho=\rho_0$ where $\rho_0$ is a constant, in general we get a flat de Sitter solution. 
%For other cosmological bounce solutions see Refs. \cite{Biswas1,Biswas_2} %and references therein.

\subsubsection{Modified Born-Infeld model}

In the literature several other modified teleparallel gravity bounce models have been proposed. In this section we revisit one of them. In particular, we study a Born-Infeld type model based on the choice suggested in \cite{FerraroBI}, and also discussed in \cite{Sadjadi:2012xa}.  For other models see \cite{Topo20}
\begin{equation}
   f(T)=\frac{12}{\alpha^2}\left[\sqrt{1-\alpha^2 \frac{T}{6}}-1
 \right]\,,  \label{mod2}
\end{equation}
which can be rewritten as
\begin{equation}
 f(H)=\frac{12}{\alpha^2}\left[\sqrt{1-\alpha^2 H^2}-1
 \right]\,.\label{eqbi}
\end{equation}
The related equations of motion lead to
\begin{equation}
 \left(1+\frac{\alpha^2 \rho}{6}\right)\left[\sqrt{1-\alpha^2 H^2}
 \right]=1\,,\label{eqbi2}
\end{equation}
from which we can compute the modified Friedmann equation
\begin{equation}
3H^2=\rho\left[1+\frac{\alpha^2\rho}{12}\right]\left( 1+\frac{\alpha^2\rho}{6} \right)^{-2}  \,.
\label{zf1}
\end{equation}
From the matter conservation law and a barotropic fluid $p=\omega \rho$, we obtain
\begin{equation}
\dot{\rho}=-\sqrt{3}(1+\omega)\left[\rho^3+\frac{\alpha^2\rho^4}{12}\right]^{1/2}
\left(1+\frac{\alpha^2\rho}{6} \right)^{-1}  \,,
\label{zf12}
\end{equation}
whose solution is
\begin{equation}
\sqrt{3\rho}(1+\omega)t=\left[1+\frac{\alpha^2\rho}{12}\right]^{1/2} 
-\alpha \,\sqrt{2 \rho} \, \text{arcsinh} \left[\frac{\alpha \sqrt{\rho/2}}{3}\right]\,.
\label{zf3}
\end{equation}
This equation above is a transcendental equation for $\rho$ which in general is difficult to solve analytically. However, we may find an approximated solution for small values of $|\alpha|$,
\be
\rho(t)=\frac{1}{3(1+\omega)^2 t+\frac{\alpha^2}{8}}+O(\alpha^4)\,.
\ee
Therefore also in this model we can remove the GR Big bang singularity at $t=0$, which can be retrieved in the limit $|\alpha| \rightarrow 0$.

\subsection{Cosmological perturbations}

In this section we recall the linear cosmological perturbations in the context of $f(T)$- modified gravity, in a flat FLRW space-time, in order to assess the stability of our model in the context of linear perturbations. We will use the derivation found in Ref \cite{FT43}. By introducing the comoving coordinates with conformal time $d\eta = dt / a(t)$, such that $a \equiv a(\eta)$, the choice for the unperturbed vierbein reads:
\be
e_{\mu}^A = a(\eta) \delta_{\mu}^A\,.
\ee
The torsion scalar is given by,
\be
T= \frac{6}{a^2}\mathcal  H^2\,,
\ee
where we define $\mathcal H=a'/a\equiv H a$. In what follows, the prime index will denote the derivative with respect the conformal time $\eta$.
The general perturbed vierbein at linear order reads:
\be 
e^{0}_{0} = a(\eta)(1+\phi)\,,
\ee
\be
e^{0}_{i} = a(\eta)(\p _{i}\beta + u_i)\,,
\ee
\be
e^{a}_{0} = a(\eta)(\p _{a}\zeta + v_a) \,,
\ee
\be
e_{j}^{a} = a(\eta) \left((1-\psi)\delta _{j}^{a} + \p ^2_{aj}\sigma + \epsilon_{ajk}\p _{k}s + \p _{j}c_{a} + \epsilon_{ajk}w_{k} + \frac{1}{2}h_{aj}\right)\,.
\ee
As usual, the vector perturbations are taken to be divergenceless, and the tensor perturbations traceless. 

One can immediately see that in this case, compared to GR, one has to deal with 6 new components given by the scalar $\beta + \zeta$, pseudoscalar $s$, vector $u_i + v_i$ and pseudovector $w_j$, which have to be taken into account since they correspond to the 6 degrees of freedom associated to local Lorentz rotations of the tetrad.\\

Thus, the perturbed metric elements read,
\be
g_{00} = -a^2(\eta) (1+2\phi)\,,
\ee
\be
g_{0i} = a^2(\eta) (\p _i (\zeta - \beta) + v_i -u_i)\,,
\ee
\be
g_{ij} = a^2(\eta)((1-2\psi)\delta_{ij} + 2\p ^2_{ij}\sigma + \p _{i}c_j + \p _{j}c_i + h_{ij})\,.
\ee
In the following we consider only the scalar and the tensor linear perturbations. 

\subsubsection{Scalar perturbations}

In order to study the scalar perturbations we will use the Newtonian gauge, obtained by choosing $\beta = \zeta$ and $\sigma = 0$. Therefore, one can find the linearized torsion components, and write down the variation of the torsion scalar as,
\be
\delta T = -\frac{4\mathcal H}{a^2}(\nabla ^2 \zeta + 3\mathcal H\phi + 3\psi ')\,.
\label{varscalar}
\ee
From the antisymmetric part of the equations of motion, it is possible to derive the following relation,
\be
\nabla ^2\zeta = -3\left(\psi ' + \mathcal H\phi - \frac{\mathcal H'-\mathcal H^2}{\mathcal H}\psi\right)\,,
\label{zetaconstr}
\ee
such that one easily 
see that $\zeta$ is constrained, and is not a dynamical variable.

The symmetric part of the linearized scalar perturbation equations result to be,
\be
f_T(\nabla ^2 \psi - 3\mathcal H(\psi ' + \mathcal H\phi)) - \frac{36f_{TT}\mathcal H^2(\mathcal H'-\mathcal H^2)}{a^2}\psi = \frac{1}{2} a^2 \delta\rho\,,
\label{00}
\ee

\be
f_T (\psi ' + \mathcal H\phi) + \frac{12\mathcal H(\mathcal H'-\mathcal H^2)f_{TT}}{a^2}\psi = \frac{1}{2} a^2 (\rho + p)\delta u\,,
\label{0i}
\ee

\be
f_T(\phi - \psi) + \frac{12f_{TT}\mathcal H(\mathcal H'-\mathcal H^2)}{a^2}\zeta = \delta s\,,
\label{ij}
\ee

\be\begin{split}
&f_T\left[\psi '' + \mathcal H(2\psi' + \phi') + (\mathcal H^2+2\mathcal H')\phi + \frac{1}{3}\nabla^2(\phi-\psi)\right] +\\ 
&+\frac{12f_{TT}}{a^2}\left[\mathcal H(\mathcal H'-\mathcal H^2)\psi ' + (\mathcal H\mathcal H'' + 2\mathcal H'^2 - 5\mathcal H^2 \mathcal H' + \mathcal H^4)\psi + \mathcal H^2(\mathcal H'-\mathcal H^2)\phi\right] +\\
&+\frac{144f_{TTT}\mathcal H^2(\mathcal H'-\mathcal H^2)^2\psi}{a^4} = \frac{1}{2} a^2\delta p\,,
\label{ii}
\end{split}\ee
where it has made use of (\ref{zetaconstr}) and the functions $\delta \rho$, $\delta p$, $\delta u$ and $\delta s$ are the fluctuations of energy density, pressure, fluid velocity and anisotropic stress, respectively.

Note that even by assuming a vanishing anisotropic stress, $\delta s=0$, one deals with a gravitational slip $\phi-\psi\neq 0$ when $f_{TT}\neq 0$.
Recalling that for adiabatic perturbations it is possible to relate the energy density and pressure, namely $\delta p = c_s^2 \delta \rho$, we can derive from the system above with $\delta s=0$ the following relation,
\begin{eqnarray}
&&f_T\left(
\psi''+\mathcal H (2\psi'+\mathcal H\phi')+\phi(\mathcal H^2+2\mathcal H')
\right)\nonumber\\&&
+\frac{12f_{TT}}{a^2}
\left(
2\mathcal H\mathcal H'\psi'-2\mathcal H^3\psi'+\mathcal H\mathcal H''\psi+\mathcal H'^2\psi-3\mathcal H'\mathcal H^2\psi+2\phi\mathcal H^2(\mathcal H'-\mathcal H^2)
\right)\nonumber\\&&
+\frac{144f_{TTT}\mathcal H^2(\mathcal H'-\mathcal H^2)^2\psi}{a^4}
-
c_s^2f_T(\nabla^2\psi)=-3\mathcal H(\rho+p)c_s^2\delta u\,.\label{Master}
\end{eqnarray}

Moreover, one can use
Eq. (\ref{zetaconstr}) and Eq. (\ref{ij}) with $\delta s=0$ in order to find for a given wavenumber $k$,
\begin{equation}
 \phi= \frac{36f_{TT}(\mathcal H^2-\mathcal H')(\mathcal H\psi'-(\mathcal H'-\mathcal H^2)\psi)+k^2a^2f_T\psi}{k^2 a^2 f_T+36f_{TT}(\mathcal H'-\mathcal H^2)\mathcal H^2}\,.  
\end{equation}
By inserting this expression in (\ref{Master}) and in the limit $k\rightarrow\infty$ the gravitational slip vanishes and we get,
\begin{align}
\psi''&+ 3 \mathcal{H} \left[1-\frac{8}{a^2 f_T}\left(f_{TT} \mathcal{H}^2 - f_{TTT}\mathcal{H}'\right)\right]\psi' +  \nonumber\\
&+ \left[\mathcal{H}^2 + c_s^2 k^2 + \dots \right]\psi=-3\mathcal H(\rho+p)c_s^2\delta u\,,
\end{align}
where the dots stands for additional terms not depending on $k^2$ or of higher order in the $1/k$ expansion. Therefore the propagation velocity of scalar perturbations corresponds to the standard matter sound speed $c_s$ as in GR. On the other hand, for small values of $k$, the gravitational slip may diverge (see equations \eqref{ij} and \eqref{zetaconstr}). 
In particular, for small values of $|\alpha|$, a term proportional to $\alpha^2/k^2$ appears in the models (\ref{mod1}) and (\ref{mod2}). However, this behaviour is not an issue as the matter density perturbations are not divergent, see Ref. \cite{FT43}. And also in the large scales case the $c_s^2$ is not modified with respect to General Relativity. Therefore we expect our models not to be affected by gradient instabilities in the late universe.

\subsubsection{Tensor perturbations}
It is easy to see that tensor perturbations do not affect the variation of the torsion scalar, while the
equation for the tensor perturbations reads:
\be
\left(h''_{ij} + 2\mathcal H h'_{ij} - \nabla ^2h_{ij}\right) + \frac{12(\mathcal H'-\mathcal H^2)f_{TT}}{a^2 f_T} h'_{ij} = 0\,.
\ee
Although we have a new friction term, there are no new mass terms. We can therefore conclude that in general $f(T)$ theories do not introduce massive gravitons. Or, in other words, the velocity of tensor waves is equal to the speed of light at any time. Finally, the limit $\alpha\rightarrow 0$ recovers the behaviour of GR perturbations at linear order.\\

\subsection{Non-flat FLRW case}

In this section we review the teleparallel gravity in a non spatially flat FLRW space-time. In fact, the generalization of what we previously obtained in the flat to non flat FLRW case is not straightforward, due to the subtleties associated with the evaluation of the torsion scalar $T$. Once $T$ is computed, since we are considering a dynamical highly spherical symmetric space-time, we may apply the superspace methods (see for example Ref. \cite{Monica}).  For a more general approach, see for example \cite{H2} and reference therein. 

Consider the spatially curved FLRW, written in spherical coordinates
\be
ds^2=-dt^2+a(t)^2\left(\frac{dr^2}{1-kr^2}+r^2 d\Omega_2^2  \right)\,,
\label{metricn}
\ee
where we  units such that the curvature scalar $|k|=1$.
The key point is the evaluation of the scalar tensor $T$ on this space-time. In a non spatially flat FLRW in spherical coordinates, the naive choice for the tetrad requires a suitable non vanishing spin connection, or alternatively, in the gauge $\omega=0$,  one has to consider a non trivial suitable choice for the tetrad (see for example Refs. \cite{Ferraro11, H1,Capo20}). In both approaches, the result is the same and reads
\be
T=6\left[H^2(t)-\frac{k}{a^2(t)}\right]\,.
\label{c}
\ee
However, this is not the unique expression for the scalar $T$ in a spatially curved space-time. In fact, according to \cite{H20}, for $k<0$ we may also have
\be
T=6\left[H(t) \pm \frac{1}{a(t)}\right]^2\,.
\label{h1}
\ee

In all cases, applying the superspace method, the action can be written, up to a trivial multiplicative time independent factor, as
\be
I=\int a^3 f[T(a, \dot{a})]dt+I_m\,,
\ee
where $I_m$ is the usual matter Lagrangian.

Firstly we deal with the choice of the torsion scalar (\ref{c}). We can consider $a(t)$ as Lagrangian coordinate, and from the variation of the action with respect to it we obtain the second Friedmann equation
\be
-p=\frac{f(T)}{2}-f_T\left( 6H^2+2\dot{H}-2\frac{k}{a^2}\right)-24H^2\left(\dot{H}+\frac{k}{a^2}\right)f_{TT}\,.
\ee
The first Friedmann equation follows from the above equation and the standard matter conservation law, ensured by the diffeomorphism invariance of the model, and reads
\be
\rho=\frac{f(T)}{2}-6H^2f_T\,.
\label{kno2}
\ee
For the second choice of the torsion scalar(\ref{h1}), the second Friedmann equation reads
\be
-p=\frac{f(T)}{2}-2f_T\left( 3H^2+\dot{H}+2\frac{1}{a^2} \pm 3\frac{H}{a}       \right)-24\left(H \pm \frac{1}{a} \right)^2\left(\dot{H} \mp \frac{H}{a^2}\right)f_{TT}\,,
\label{h2}
\ee
while the first Friedmann equation is
\be
\rho=\frac{f(T)}{2}-6H \left( H \pm \frac{1}{a} \right) f_T\,,
\ee
in agreement with the results of \cite{H20}. As a simple consistency check, in the GR case $f(T)=-T$, and it both cases we obtain the GR result
\be
3H^2+\frac{3k}{a^2}=\rho\,.
\label{fgr}
\ee

In the next sections we will consider two applications of teleparallel gravity in the non flat case. In particular, we will study the models (\ref{mod1}) and (\ref{mod2}) in the non-flat case.

\subsubsection{Example 1}

Consider again the expression used in (\ref{mod1}). In this example we consider only the first class of models, where $T$ is given by (\ref{c}). The other choice leads to similar results. From Eq. (\ref{kno2}) we obtain
\begin{equation}
\frac{6}{\alpha^2}
\left(
1-\sqrt{1-\alpha^2\frac{T}{6}}
\right)=\rho-\frac{6k}{a^2}
\left[
\frac{1}{\alpha\sqrt{ T/6}}\arcsin
\left(\alpha \sqrt{\frac{T}{6}}\right)
\right]\,,\quad T=6\left(H^2-\frac{k}{a^2}\right)\,,
\end{equation}
which can be rewritten as
\begin{equation}
3\left(H^2-\frac{k}{a^2}\right)=
\rho-\frac{6k}{a^2}\psi_k(T)-\frac{\alpha^2}{12}\left[\rho-\frac{6k}{a^2}\,\psi_k(T)\right]^2\,,
\end{equation}
where
\begin{equation}
\psi_k(T)\equiv\frac{1}{\alpha\sqrt{T/6}}\arcsin\left(\alpha\frac{T}{6}\right)\,.    
\end{equation}
At the first order in $\alpha^2$ the equation above become
\begin{equation}
 3\left(H^2-\frac{k}{a^2}\right)=\left(
 \rho-\frac{6k}{a^2}
 \right)\left(1-
 \frac{\rho-\frac{6k}{a^2}}{\rho_c}
 \right)\,,\quad
 \rho_c=\frac{12}{\alpha^2}\,,
\end{equation}
or
\begin{equation}
 3\left(H^2+\frac{k}{a^2}\right)=
 \rho-
 \frac{\left(\rho-\frac{6k}{a^2}\right)^2}{\rho_c}\,,\quad
 \rho_c=\frac{12}{\alpha^2}\,.\label{superseba}
\end{equation}
The case is analog to the one discussed in \cite{nostri2}, where the Big-Bang singularity at $t=0$ is absent.
An exact solution can be found if we take a barotropic equation of state  $p=-\rho/3$, namely $\omega=-1/3$ and $\rho(t)=\rho_0 a(t)^{-2}$. 
We can rewrite equation the above equation introducing  a new variable $y= a(t)^2$
\begin{equation}
\frac{3}{4} \dot{y}^2=(\rho_0-3k)y -\frac{(\rho_0-6k)^2}{\rho_c} \,,
\quad y=a^2(t)\,,
\end{equation}
whose solution is
\be
y(t)\equiv a^2(t)=\frac{(\rho_0-6k)^2}{\rho_c(\rho_0-3k)}+\frac{\rho_0-3k}{3}t^2\,,
\ee
where we assume $\rho_0>3k$. This (bounce) solution is regular at $a(0) \neq 0$. When $\rho_c$ goes to infinity, we recover the GR solution with the Big Bang singularity.

Moreover, for the case of a generic constant value of $\omega$, we can prove the absence of singularities.
In fact, considering a perfect fluid with $p=\omega\rho$ and $\rho=\rho_0 a^{-3(1+\omega)}$, with a constant $\rho_0$, we can rewrite Eq. (\ref{superseba}) as
\begin{equation}
  \int \frac{dy}{\sqrt{Y(y)}}=t\,,\quad y=a(t)^2\,,  
\end{equation}
where $Y(y)$ is
\be
Y(y)=\frac{4}{3}\left(\rho_0 y^{\frac{1-3\omega}{2}}
-3ky-\frac{\rho_0^2y^{-1-3\omega}+36k^2-12k\rho_o y^{\frac{-1-3\omega}{2}}}{\rho_c}
\right)\,.
\ee
We can find an approximate solution, valid for small $t$ around the critical point $Y(y_*)=0$
\begin{equation}
y(t)\simeq y_*+\frac{Y'_* t^2}{4}\,,    
\end{equation}
where $t_*$ is the solution of the transcendental equation 
\be
\rho_0 y_*\left( y_*^{\frac{-1-3\omega}{2}}
-\frac{3k}{\rho_0}\right)-\frac{\rho_0^2y_*^{-1-3\omega}+36k^2-12k\rho_o y_*^{\frac{-1-3\omega}{2}}}{\rho_c}=0\,.\label{Y90}
\ee
Therefore only in the GR limit $\rho_c\rightarrow\infty$ (i.e. $\alpha\rightarrow 0$) we have a singular solution for the flat case with $k=0$, otherwise $y_*\neq 0$ and the bounce occurs, independently on the space curvature. For example, considering $k=0$, Eq. (\ref{Y90}) becomes
\be
1 -\frac{\rho_0 y_*^{\frac{-3(1+\omega)}{2}}}{\rho_c}=0\,,
\ee
whose solution reads
\begin{equation}
y_*\equiv a^2(t_*)=  \left(\frac{\rho_c}{\rho_0}\right)^{-\frac{2}{3(1+\omega)}}\,,  
\end{equation}
which corresponds to the minimum value of the scale factor on the cosmological bounce.

\subsubsection{Example 2}

Consider the model (\ref{mod2}) in non-flat FLRW space-time. Again, we use the case where the torsion scalar is given by (\ref{c}). The first Friedmann equation becomes
\begin{equation}
3\left(H^2-\frac{k}{a^2}\right)=
\frac{\rho\left(1+\frac{\alpha^2\rho}{12}\right)-\frac{6k}{a^2}-\frac{3\alpha^2k^2}{a^4}}{\left(\frac{\alpha^2\rho}{6}+1\right)^2}\,.
\end{equation}
For a perfect fluid we can write
\begin{equation}
  \int \frac{dy}{\sqrt{Y(y)}}=t\,,\quad y=a(t)^2\,,  
\end{equation}
where, at the first order of $\alpha^2$, $Y(y)$ is
\begin{equation}
Y(y)=
\frac{2}{27} \alpha ^2 \rho_0^2 (3 k-5 y) y^{-3 \omega -2}+\frac{8}{3} \alpha
   ^2 k \rho_0 y^{-\frac{3 \omega }{2}-\frac{1}{2}}-4 k \left(\alpha ^2
   k+y\right)+\frac{4}{3} \rho_0 y^{\frac{1}{2}-\frac{3 \omega }{2}}
\end{equation}
Therefore an approximate solution, valid for small $t$ around the critical point $Y(y_*)=0$, can be found and reads
\begin{equation}
y(t)\simeq y_*+\frac{Y'_* t^2}{4}\,,    
\end{equation}
where $y_*$ is the solution of 
\begin{equation}
\frac{2}{27} \alpha ^2 \rho_0^2 (3 k-5 y_*) y_*^{-3 \omega -2}+\frac{8}{3} \alpha
   ^2 k \rho_0 y_*^{-\frac{3 \omega }{2}-\frac{1}{2}}-4 k \left(\alpha ^2
   k+y_*\right)+\frac{4}{3} \rho_0 y_*^{\frac{1}{2}-\frac{3 \omega }{2}}=0\,.
\end{equation}
Thus in the limit $\alpha\rightarrow 0$ and $k=0$ we recover the Big bang singularity at $y_*=0$, otherwise the bounce appears.

\section{Conclusions}

In this paper we have investigated a derivation of Friedmann equations of QLC-like cosmology in the framework of modified teleparallel gravity, which offers a very interesting alternative approach with respect to modified gravitational theories based on metric formulation as $F(R)$ modified models. In fact, instead of using the Levi-Civita connection, teleparallel gravity is based on the so called Weitzenbock connection with a related  vanishing curvature tensor, but non vanishing torsion tensor. All the GR results can be found in the equivalent teleparallel GR formulation. Furthermore, the extension of the theory to modified teleparallel gravity preserves the field equations at the second order: this is one of the reasons why modified teleparallel gravity is often used to investigate a wide variety of solutions in a cosmological context. 

We proposed a model, studying it firstly in flat FLRW space-time, which features a non linear correction to GR equations that depends on a critical density and it is sufficient to avoid the Big Bang singularity, namely the model admits the cosmological bounce of QLC cosmology. We also argued that we can recover the GR limit, when the critical density goes to infinity. Moreover the analysis of scalar and tensor perturbations shows the viability of the theory and the lack of superluminar effects.

We also generalized the results to the curved FLRW space-time. Even though the non-flat case is more complex due to the nature of teleparallel theories, a direct evaluation of the field equations from the on-shell form of the Lagrangian, once the torsion scalar has been determined, is easily achieved. This is possible because we are working on highly symmetric space-times, namely FLRW space-times. Despite the fact that exact solutions can be found only for some specific matter choices, we investigated approximate solutions near $t=0$, and  proved, using the critical points of the theory that the models do not contain singularities and provide  cosmological bounce solutions. Finally, we have generalized the Born-Infeld model to spatially FLRW curved space.

A last remark is in order. As we pointed out in the introduction, several modified gravity models lead to loop modified equation for flat FLRW space-time.
This sort of degeneracy may be removed making use of the concept of gravitational standard candles \cite{GasperiniMaggiore}. With regard to this issue it seems crucial to deal with models having $c_T=1$, but frictional term (or effective Plank mass) in the tensor perturbation equation different from the GR one. We point out that our $F(T)$-models may be included in this list.

Finally, in the context of LQC, it would be interesting to explore the application of $f(T)$-gravity as a low-energy effective field theory, with LQC as the underlying UV theory.

\section*{Acknowledgements}

We acknowledge the use of the \texttt{xAct} package \cite{xAct}. We thank the referee for the useful comments on the first version of this paper.

\end{document}